\documentstyle[11pt]{article}
\catcode`\@=11

\@addtoreset{equation}{section}
\def\theequation{\arabic{section}.\arabic{equation}}

\catcode`\@=11

\def\section{\@startsection{section}{1}{\z@}{3.5ex plus 1ex minus
   .2ex}{2.3ex plus .2ex}{\large\bf}}

\def\eqnarray{\let\@currentlabel=\theequation\refstepcounter{equation}
    \global\@eqnswtrue
    \global\@eqcnt\z@\tabskip\@centering\let\\=\@eqncr
    $$\halign to \displaywidth\bgroup\@eqnsel\hskip\@centering
      $\displaystyle\tabskip\z@{##}$&\global\@eqcnt\@ne
       \hfil${{}##{}}$\hfil
      &\global\@eqcnt\tw@ $\displaystyle\tabskip\z@{##}$\hfil
       \tabskip\@centering&\llap{##}\tabskip\z@\cr}
\def\lefteqn#1{\hbox to 4\arraycolsep{$\displaystyle #1$\hss}}
\def\thesection{\arabic{section}.}

\def\appendix{\setcounter{section}{0}
        \def\thesection{Appendix.}
        \def\theequation{\Alph{section}.\arabic{equation}}}

\long\def\@makefntext#1{\parindent 0cm\noindent
\hbox to 1em{\hss$^{\@thefnmark}$}#1}
\def\IR{{\hbox{{\rm I}\kern-.2em\hbox{\rm R}}}}
\def\IH{{\hbox{{\rm I}\kern-.2em\hbox{\rm H}}}}
\def\IC{{\ \hbox{{\rm I}\kern-.6em\hbox{\bf C}}}}
\def\IZ{{\hbox{{\rm Z}\kern-.4em\hbox{\rm Z}}}}
\def\rref#1{(\ref{#1})}
\newcommand{\beq}{\begin{equation}}
\newcommand{\eeq}{\end{equation}}
\newcommand{\vol}{{\tilde v}}
\begin{document}
%
%
%
%
\def\citen#1{%
\edef\@tempa{\@ignspaftercomma,#1, \@end, }
\edef\@tempa{\expandafter\@ignendcommas\@tempa\@end}%
\if@filesw \immediate \write \@auxout {\string \citation {\@tempa}}\fi
\@tempcntb\m@ne \let\@h@ld\relax \let\@citea\@empty
\@for \@citeb:=\@tempa\do {\@cmpresscites}%
\@h@ld}
%
\def\@ignspaftercomma#1, {\ifx\@end#1\@empty\else
   #1,\expandafter\@ignspaftercomma\fi}
\def\@ignendcommas,#1,\@end{#1}
%
%
\def\@cmpresscites{%
 \expandafter\let \expandafter\@B@citeB \csname b@\@citeb \endcsname
 \ifx\@B@citeB\relax 
    \@h@ld\@citea\@tempcntb\m@ne{\bf ?}%
    \@warning {Citation `\@citeb ' on page \thepage \space undefined}%
 \else
    \@tempcnta\@tempcntb \advance\@tempcnta\@ne
    \setbox\z@\hbox\bgroup 
    \ifnum\z@<0\@B@citeB \relax
       \egroup \@tempcntb\@B@citeB \relax
       \else \egroup \@tempcntb\m@ne \fi
    \ifnum\@tempcnta=\@tempcntb 
       \ifx\@h@ld\relax 
          \edef \@h@ld{\@citea\@B@citeB}%
       \else 
          \edef\@h@ld{\hbox{--}\penalty\@highpenalty \@B@citeB}%
       \fi
    \else   
       \@h@ld \@citea \@B@citeB \let\@h@ld\relax
 \fi\fi%
 \let\@citea\@citepunct
}
%
\def\@citepunct{,\penalty\@highpenalty\hskip.13em plus.1em minus.1em}%
%
%
\def\@citex[#1]#2{\@cite{\citen{#2}}{#1}}%
%
%
\def\@cite#1#2{\leavevmode\unskip
  \ifnum\lastpenalty=\z@ \penalty\@highpenalty \fi 
  \ [{\multiply\@highpenalty 3 #1
      \if@tempswa,\penalty\@highpenalty\ #2\fi 
    }]\spacefactor\@m}
\let\nocitecount\relax  
%
\begin{titlepage}
\vspace{.5in}
\begin{flushright}
UCD-97-19\\
August 1997\\
revised September 1997\\
gr-qc/9708026\\
\end{flushright}
\vspace{.5in}
\begin{center}
{\Large\bf
 Spacetime Foam\\[1ex] and the Cosmological Constant}\\
\vspace{.4in}
{S.~C{\sc arlip}\footnote{\it email: carlip@dirac.ucdavis.edu}\\
       {\small\it Department of Physics}\\
       {\small\it University of California}\\
       {\small\it Davis, CA 95616}\\{\small\it USA}}
\end{center}

\vspace{.5in}
\begin{center}
\begin{minipage}{4.2in}
\begin{center}
{\large\bf Abstract}
\end{center}
{\small
In the saddle point approximation, the Euclidean path integral
for quantum gravity closely resembles a thermodynamic partition
function, with the cosmological constant $\Lambda$ playing the
role of temperature and the ``density of topologies'' acting as
an effective density of states.  For $\Lambda<0$, the density of
topologies grows superexponentially, and the sum over topologies
diverges.  In thermodynamics, such a divergence can signal the
existence of a maximum temperature.  The same may be true in
quantum gravity: the effective cosmological constant may be
driven to zero by a rapid rise in the density of topologies.
}
\begin{flushleft}
\end{flushleft}
\end{minipage}
\end{center}
\end{titlepage}
\addtocounter{footnote}{-1}

The cosmological constant $\Lambda$---in modern language, the energy
density of the vacuum---is observed to be less than $10^{-47}
\mathrm{GeV}^4$, or $10^{-120}$ in Planck units.  The cosmological
constant problem \cite{Weinberg,Ng}, the problem of explaining the
smallness of this number, is one of the central puzzles of modern
physics.  A  natural guess is that some symmetry forces $\Lambda$ to
vanish, but the two obvious candidates, supersymmetry and conformal
symmetry, are both badly broken.  One can, of course, set $\Lambda$
to zero by fiat, but this requires fine-tuning over a vast range of
energies, and is in any case time-dependent, since phase transitions
in the early universe can change the value of $\Lambda$.  One can search
for dynamical mechanisms to relax the cosmological constant to zero,
but such attempts typically involve the implicit use of conformal
invariance, and fail when the symmetry is broken \cite{Weinberg}.

This leaves quantum gravity as a tempting place to look for an explanation.
Perhaps the most intriguing proposal to date has been Coleman's
wormhole model \cite{Coleman}, in which topological fluctuations of
spacetime induce effective nonlocal interactions that smear $\Lambda$
into a probabilistic distribution peaked sharply at zero.  The proposal
presented in this paper is similar in spirit to Coleman's, but different
in detail: I consider a different set of topologies, with metrics that
(unlike Coleman's) are exact saddle points of the functional integral,
and I interpret the resulting partition function rather differently.  In
particular, I will argue that a rapidly growing density of topologies may
drive the cosmological constant to zero, as processes that could increase
$|\Lambda|$ instead merely produce more complicated ``spacetime foam.''

\section{The Euclidean Gravitational Partition Function}

I shall work in Euclidean quantum gravity, that is, quantum gravity
``analytically continued'' to Riemannian (positive-definite) metrics,
since this seems to be the most natural setting in which to consider
fluctuations of spacetime topology.  The partition function for
the volume canonical ensemble is \cite{Hawking,Gibbons}
\beq
Z[\Lambda] = \sum_M \int [dg] \exp\{-I_E\} ,
\label{a1}
\eeq
where the sum is over topologically distinct manifolds and the
Euclidean action $I_E$ is
\beq
I_E = - {1\over16\pi L_P{}^2} \int_M (R-2\Lambda) \sqrt{g}d^4x .
\label{a2}
\eeq
($L_P$ is the Planck length.)  General relativity is not renormalizable,
so the meaning of the path integral is not entirely clear, but \rref{a2}
can be regarded as an effective action for distances much larger than
the Planck scale.

Extrema of the action \rref{a2} are Einstein metrics, with classical
actions
\beq
\bar I_E(M) = -{\Lambda\over8\pi L_P{}^2}{\mathit{Vol}}(M)
            = - {9\over8\pi\Lambda L_P{}^2}\vol(M) ,
\label{a3}
\eeq
where $\vol(M)$ is the normalized volume, obtained by rescaling the
metric to set the scalar curvature to $\pm12$.  (The factor of $12$ is
conventional; hyperbolic four-manifolds, i.e., manifolds of constant
curvature $-1$, have scalar curvature $-12$.)  Although $\vol$ is a
geometric quantity, normalized volumes of Einstein metrics characterize
topology as well.  In particular, for $\Lambda<0$ there is no known
example of a manifold that admits two Einstein metrics with different
values of $\vol$ \cite{Besse}.  Roughly speaking, $\vol(M)$ measures
the topological complexity of $M$; for a hyperbolic manifold, for
instance, $\vol(M) = 4\pi^2\chi(M)/3$, where $\chi$ is the Euler number.

In the saddle point approximation, the partition function \rref{a1} is
\beq
Z[\Lambda] = \sum_M \Delta_M \exp
  \left\{ {9\over8\pi\Lambda L_P{}^2}\vol(M) \right\} .
\label{a4}
\eeq
The prefactors $\Delta_M$ are combinations of determinants coming from
gauge-fixing and from small fluctuations around the extrema.  Their
precise values are not known, but their dependence on $\Lambda$ can be
computed from the trace anomaly \cite{Christensen}: up to a possible 
polynomial dependence coming from zero-modes,
\beq
\Delta_M \sim \Lambda^{-\gamma/2} , \qquad
\gamma = {106\over45}\chi(M) - {261\over40\pi^2}\vol(M) .
\label{a5}
\eeq
For our purposes, the crucial observation is that $\Delta_M$ is no more 
than exponential in $\vol$.

We shall be primarily interested in manifolds with $\Lambda<0$; this
is typical for most topologies \cite{Hawking}.  We can thus rewrite
equation \rref{a4} as
\beq
Z[\Lambda] = \sum_\vol \rho(\vol) \exp
  \left\{ -{9\over8\pi|\Lambda| L_P{}^2}\vol \right\} ,
\label{a6}
\eeq
where $\rho(\vol)$ is a ``density of topologies'' that counts the
number of manifolds (weighted by $\Delta_M$) with a given value of
$\vol$.

Equation \rref{a6} closely resembles the expression for the canonical
partition function of a thermodynamic system,
\beq
Z_{\mathrm{thermo}}[\beta] = \sum_E \rho(E)\exp\{ -\beta E \} ,
\label{a7}
\eeq
where the ``temperature'' for the gravitational partition function is
$\beta^{-1} = 8\pi|\Lambda|L_P{}^2/9$.  The analogy is not exact, of
course: the gravitational partition function does not describe dynamics
(it is already four dimensional!), so there is no obvious equivalent of
heat flow.  But the correspondence goes beyond the formal similarity
of equations \rref{a6} and \rref{a7}.  Like the energy in a thermodynamic
system, the normalized volume $\vol(M)$ can be divided among small regions
of $M$, with weak interactions coming from the need to add boundary terms
to the action for an open region.  Moreover, even without a dynamical
model of topology change in which to derive an ergodic theorem, we know
that by construction, manifolds with the same value of $\vol$ occur with
equal probabilities (up to loop corrections).

Until now, the standard assumption in Euclidean quantum gravity has
been that $\rho(\vol)$ grows no faster than polynomially in $\vol$.  As
we shall see below, this assumption is incorrect.  To understand the
significance of of this observation, let us first consider the
thermodynamic analog.

\section{Thermodynamics with a Rapidly Growing Density of States}

The thermodynamics of a system with an exponentially growing density of
states was first considered by Hagedorn in the context of the hadron
mass spectrum in bootstrap models \cite{Hagedorn,Hagedorn2}.  Suppose
$\rho(E)$ takes the form
\beq
\rho(E) = E^a e^{bE} .
\label{b1}
\eeq
The sum \rref{a7} then converges only for $\beta>b$.  The Hagedorn
temperature $T=1/b$ is a maximum temperature: as $T$ approaches $1/b$,
the expectation value of the energy diverges, as does the heat capacity.
While this phenomenon may be surprising, its physical explanation is
fairly simple.  Energy added to a system can either go into increasing
the energy of existing states or into creating new states.  If the density
of states rises rapidly enough, many more new states are available at
higher energies; as the temperature approaches its critical value, added
energy goes entirely toward creating new states rather than heating those
already present.

If $\rho(E)$ grows faster than exponentially, the partition function
\rref{a7} has a vanishing radius of convergence, and the maximum
temperature effectively shrinks to zero \cite{Hagedorn2}.  To investigate
a system of this sort, one must use the microcanonical ensemble.  The
microcanonical inverse temperature is
\beq
\beta = {\partial \ln\rho(E)\over\partial E} ,
\label{b2}
\eeq
and the heat capacity is
\beq
c_V = -\beta^2\left({\partial^2 \ln\rho(E)\over\partial E^2} \right)^{-1} .
\label{b3}
\eeq
The condition that the density of states rise superexponentially is
precisely that the second derivative in \rref{b3} be positive, and that
$c_V$ thus be negative.

Systems with negative heat capacities have been studied by a number of
authors \cite{Thirring,Thirring2,Lynden,Lynden2,Landsberg}.  Such systems
are thermodynamically unstable; placed in contact with a heat bath, they
will experience runaway heating or cooling.  Nevertheless, they can
occur in nature, and it is possible to make sense of their thermodynamic
properties.  In particular, $\beta^{-1}$ should now be understood as the
temperature measured by a small thermometer rather than a large heat bath
\cite{Thirring}.  This quantity retains much of its usual statistical
significance: if one starts with a large system with fixed energy
$\bar E$ and considers small subsystems with energies $E\ll\bar E$,
the probability of finding a given energy $E$ is proportional to
$\exp\{-\beta E\}$.  Unlike ordinary thermodynamic systems, however,
a system with negative heat capacity does not distribute its energy
evenly among subsystems; the most probable configurations are those in
which almost all of the energy is concentrated in a single subsystem.

Systems with maximum temperatures and those with negative heat capacities
occur in rather different contexts, but their thermodynamic behavior has
a common physical basis.  If the density of states grows exponentially, an
inflow of energy at the Hagedorn temperature goes entirely into producing
new states, leaving the temperature constant.  If the density of states
grows superexponentially, the process is similar, but the production of
new states is so copious that an inflow of energy actually drives the
temperature down.

\section{The Density of Topologies}

The question now before us is how fast the ``density of topologies''
$\rho(\vol)$ in \rref{a6} grows as $\vol$ increases.  The full answer is
not known, but some recent mathematical results make it possible to show
that the growth is superexponential.

In particular, a lower bound can be found by considering hyperbolic
metrics, which are, of course, automatically Einstein metrics.  If $M$
is a hyperbolic manifold with normalized volume $\vol$, any $n$-fold
covering of $M$ is a hyperbolic manifold with volume $n\vol$.  Covering
spaces come from subgroups of the fundamental group $\pi_1(M)$---a subgroup
of index $n$ gives an $n$-fold cover---so if the number of index-$n$
subgroups can be estimated, this will give us partial information about
the number of hyperbolic manifolds.

Lubotzky has recently demonstrated that for a large class of hyperbolic
manifolds, $\pi_1(M)$ has a finite-index subgroup that maps homomorphically
onto a nonabelian free group $F_k$ \cite{Lubotzky}.  Such a map allows us
to construct a subgroup of $\pi_1(M)$ for each subgroup of $F_k$.  But the
number of index-$n$ subgroups of $F_k$ is known to grow  asymptotically
as $(n!)^{k-1}$ \cite{Hall}, so the number of index-$n$ subgroups of
$\pi_1(M)$ must grow at least as fast.  There is a subtlety in the next
step of the argument: while each subgroup of $\pi_1(M)$ determines a
covering space of $M$, different subgroups can sometimes give the same
covering space.  For a particular class of four-manifolds with nonarithmetic
fundamental groups, however, this overcounting can be controlled, and it
may be shown that the number of distinct covering spaces of volume $n\vol$
grows at least factorially with $n$ \cite{Lubotzky2,Carlip}.  The total
number of hyperbolic manifolds thus grows at least factorially with
normalized volume, that is,
\beq
\rho(\vol) > c_0 \exp\{ c_1\vol\ln\vol \}
\label{c1}
\eeq
for some constants $c_0$ and $c_1$.

This factorial bound probably seriously underestimates the actual growth
of $\rho(\vol)$.  Indeed, our result comes from looking only at hyperbolic
metrics---and a limited class of hyperbolic metrics, at that---and most
four-manifolds do not admit such metrics.  But the lower bound \rref{c1}
is already strong enough to guarantee that the sum over topologies
diverges, and is not even Borel summable unless higher loop
terms introduce relative phases among topologies.

Moreover, our derivation makes it clear that short-distance physics
alone cannot cure this divergence.  Indeed, the covering spaces we have
considered look alike locally, and can be distinguished only by their
long-distance properties.  The divergence comes not from high topological
complexity in small regions, but rather from the huge variety of possible
identifications of distant points in large universes.  Convergence of
the sum \rref{a6} would thus require an infrared cutoff as well as
(probably) an ultraviolet cutoff.  Actually, the existence of an IR
cutoff is not implausible: at one loop, the resummed effective action
contains nonlocal terms involving inverse Laplacians \cite{Vilkovisky},
and the eigenvalues of Laplacians typically become small when $\vol$ is
large.

A similar divergence occurs in the sum over topologies in string theory
\cite{Gross}.  In two dimensions, this divergence can be handled by
appealing to matrix models \cite{matrix}, although the cure requires that
we abandon any fundamental role for smooth geometries.  In four dimensions,
however, we know of no such solution, and must therefore ask whether any
sense can be made of the sum over topologies.

A possible answer comes from the thermodynamic analog of the preceding
section.  Let us impose an infrared cutoff---its details do not matter,
and it may ultimately be removed---to force the sum \rref{a6} to converge.
The sum will then be dominated by topologies with normalized volumes near
some maximum $\vol_{\mathrm{max}}$.  We can now consider a microcanonical
ensemble with fixed $\vol = \vol_{\mathrm{max}}$, and ask about the
expected behavior of smaller regions of a large universe.  In particular,
the ``microcanonical'' cosmological constant will be
\beq
\Lambda = - {9\over8\pi L_P{}^2}
  \left( {\partial\ln\rho(\vol)\over\partial\vol} \right)^{-1}
  \Biggl|_{\vol_{\mathrm{max}}} ,
\label{c2}
\eeq
which becomes small as $\vol_{\mathrm{max}}$ becomes large.

The rate of fall-off of $\Lambda$ depends on the exact form of $\rho(\vol)$.
It is rather slow for the factorial growth of equation \rref{c1}, but
we know this expression underestimates the true growth rate.  As in
Coleman's wormhole model \cite{Coleman}, it is plausible that this rate
will exponentiate when we take into account, for example, connected
sums of hyperbolic manifolds.  If this is the case, $\Lambda$ will be
exponentially suppressed as $\vol_{\mathrm{max}}$ increases.  The mechanism
for this suppression can be understood from the thermodynamic analogy:
rather than increasing the observed cosmological constant, an attempt
to increase $|\Lambda|$ will merely drive the production of more and more
complicated spacetime foam.

The missing element of this analysis, of course, is a detailed dynamical
picture.  An intrinsically four-dimensional formalism like the Euclidean
path integral is ill-suited for describing the temporal evolution of
$\Lambda$.  To some extent, this difficulty is inherent in quantum gravity:
it is never easy to describe dynamics in a theory with no fixed background
with which to measure the passage of time \cite{Kuchar}.  But it would
be interesting to examine the effect of the growth of $\rho(\vol)$ in
other settings, for instance in the computation of transition amplitudes
or the Hartle-Hawking wave function.

It would also be interesting to apply a similar ``thermodynamic'' analysis
to the case of a positive cosmological constant.  It is evident from
equation \rref{a4} that positive $\Lambda$ is analogous to negative
temperature.  This is consistent with the behavior of $\rho(\vol)$ for
$\Lambda>0$: $\vol$ has a maximum value of $8\pi^2/3$, the normalized
volume of a four-sphere, and the density of topologies increases as $\vol$
decreases, much as the density of states behaves in a system with a
negative spin temperature \cite{spin}.

\vspace{1.5ex}
\begin{flushleft}
\large\bf Acknowledgements
\end{flushleft}

I received help from a number of mathematicians, including Walter Carlip,
Greg Kuperberg, Alex Lubotzky, and Bill Thurston.  This work was supported
in part by National Science Foundation grant PHY-93-57203 and Department
of Energy grant DE-FG03-91ER40674.

\end{document}